\documentclass[pdflatex,sn-mathphys-num]{sn-jnl}

\usepackage{graphicx}%
\usepackage{multirow}%
\usepackage{amsmath,amssymb,amsfonts}%
\usepackage{amsthm}%
\usepackage{mathrsfs}%
\usepackage[title]{appendix}%
\usepackage{xcolor}%
\usepackage{textcomp}%
\usepackage{manyfoot}%
\usepackage{booktabs}%
\usepackage{algorithm}%
\usepackage{algorithmicx}%
\usepackage{algpseudocode}%
\usepackage{listings}%
\usepackage[acronyms]{glossaries}

\newacronym{AOM}{AOM}{acousto-optic modulator}
\newacronym{AWG}{AWG}{arbitrary waveform generator}
\newacronym{BER}{BER}{bit error rate}
\newacronym{CD}{CD}{chromatic dispersion}
\newacronym{DP}{DP}{dual-polarisation}
\newacronym{DSP}{DSP}{digital signal processing}
\newacronym{EDFA}{EDFA}{erbium-doped fiber amplifier}
\newacronym{EVM}{EVM}{error vector magnitude}

\newacronym{FDL}{FDL}{fibre delay line}
\newacronym{FSO}{FSO}{free-space optical}
\newacronym{GMI}{GMI}{generalized mutual information}
\newacronym{HD-FEC}{HD-FEC}{hard-decision forward error correction}
\newacronym{HG}{HG}{Hermite-Gaussian}
\newacronym{IQ}{IQ}{in-phase and quadrature}
\newacronym{LG}{LG}{Laguerre-Gaussian}

\newacronym{MDM}{MDM}{mode-division multiplexing}
\newacronym{MIMO}{MIMO}{multiple-input multiple-output}
\newacronym{MPLC}{MPLC}{multi-plane light conversion}
\newacronym{OAM}{OAM}{orbital angular momentum}
\newacronym{OF}{OF}{objective function}
\newacronym{OSA}{OSA}{optical spectrum analyser}
\newacronym{OSNR}{OSNR}{optical signal-to-noise ratio}
\newacronym{PRBS}{PRBS}{pseudo random binary sequence}
\newacronym{PS}{PS}{probabilistic shaping}
\newacronym{QAM}{QAM}{quadrature amplitude modulation}
\newacronym{QPSK}{QPSK}{quadrature phase-shift keying}
\newacronym{RRC}{RRC}{root-raised cosine}
\newacronym{SDM}{SDM}{space-division multiplexing}
\newacronym{SISO}{SISO}{single-input single-output}
\newacronym{SMF}{SMF}{single-mode fibre}
\newacronym{SNR}{SNR}{signal-to-noise ratio}
\newacronym{TDM}{TDM}{time-division multiplexing}
\newacronym{VOA}{VOA}{variable optical attenuator}
  \glsdisablehyper 

\usepackage[colorinlistoftodos]{todonotes}

\begin{document}

\title[Article Title]{Beaconless Auto-Alignment for Single-Wavelength 5 Tbit/s Mode-Division Multiplexing Free-Space Optical Communications}


\author*[1]{\fnm{Yiming} \sur{Li}}\email{liy70@aston.ac.uk}

\author[2]{\fnm{Gil} \sur{Fernandes}}\email{gfernandes@av.it.pt}

\author[1]{\fnm{David} \sur{Benton}}\email{d.benton@aston.ac.uk}

\author[3]{\fnm{Antonin} \sur{Billaud}}\email{antonin@cailabs.com}

\author[1]{\fnm{Mohammed} \sur{Patel}}\email{m.patel70@aston.ac.uk}

\author[1]{\fnm{Andrew} \sur{Ellis}}\email{andrew.ellis@aston.ac.uk}

\affil*[1]{\orgdiv{Aston Institute of Photonic Technologies}, \orgname{Aston University}, \orgaddress{\city{Birmingham}, \postcode{B4 7ET}, \country{UK}}}

\affil[2]{\orgdiv{Instituto de Telecomunicações}, \orgname{University of Aveiro}, \orgaddress{\city{Aveiro}, \postcode{3810-193}, \country{Portugal}}}

\affil[3]{\orgname{Cailabs}, \orgaddress{\street{1 Rue Nicolas Joseph Cugnot}, \city{Rennes}, \postcode{35000}, \country{France}}}


\abstract{Mode-division multiplexing has shown its ability to significantly increase the capacity of free-space optical communications. An accurate alignment is crucial to enable such links due to possible performance degradation induced by mode crosstalk and narrow beam divergence. Conventionally, a beacon beam is necessary for system alignment due to multiple local maximums in the mode-division multiplexed beam profile. However, the beacon beam introduces excess system complexity, power consumption, and alignment errors. Here we demonstrate a beaconless system with significantly higher alignment accuracy and faster acquisition. This system also excludes excess complexity, power consumption, and alignment errors, facilitating simplified system calibration and supporting a record-high 5.14 Tbit/s line rate in a single-wavelength free-space optical link. We anticipate our paper to be a starting point for more sophisticated alignment scenarios in future multi-Terabit mode-division multiplexing free-space optical communications for long-distance applications with a generalised mode basis.}




\maketitle

\section{Introduction}\label{sec:intro}
\Gls{MDM} is a promising technology to extend the capacity limits in \gls{FSO} communication systems \cite{wang2012terabit}. Orthogonal to other degrees of freedom such as amplitude \cite{ali201910}, phase \cite{lange2006142}, polarisation \cite{bitachon2022tbit} and wavelength \cite{ciaramella20091}, \gls{MDM} offers the spatial degree of freedom for multiplexing, extending capacity beyond the conventional limit in both \gls{FSO} and fibre communications \cite{wang2012terabit,rademacher2021peta}. Beyond conventional employment of \gls{OAM} modes \cite{wang2012terabit}, which constitute an incomplete basis, there is a growing interest in employing complete mode basis, such as \gls{HG} or \gls{LG} modes \cite{pang2018400,li2022enhanced}, to maximise potential capacity gains in the spatial domain \cite{zhao2015capacity}.

In contrast to fibre communications, \gls{FSO} communication systems require stringent alignment to maximise received power \cite{fernandes2022free}. In \gls{MDM} systems, there is an additional requirement to minimise inter-mode crosstalk \cite{abadi2019space}, thus increasing the demand for precise alignment systems. Although \gls{SISO} \gls{FSO} systems can naturally reuse the signal beam for alignment \cite{fernandes2022free}, most \gls{MDM} \gls{FSO} systems employed a beacon beam for alignment due to multiple local power extremums during the alignment \cite{li2017high,abadi2019space}. Although adaptive optics can be employed for alignment through tip/tilt compensation, it is also incompatible with generalised \gls{MDM} systems due to the presence of multiple power extremums, and a beacon beam is still required \cite{ren2014adaptive}. While attempts have been made on beaconless alignment in \gls{OAM} systems \cite{li2018experimental}, it can not support complete mode bases such as \gls{HG} or \gls{LG} modes because certain modes have zero power on the x- and y-axis. Moreover, the employed quadrant-detector-based alignment structure still introduced practical alignment errors between the quadrant detector and the signal photodiodes. The errors were mitigated by introducing order intervals, which significantly reduced the number of applicable modes. 

To minimise the inter-mode crosstalk in \gls{MDM} system, it is important to reduce the practical optical calibration error between the signal and beacon beams, as well as between the quadrant detector and the signal photodiodes. Therefore, we should eliminate any excessive \gls{FSO} paths, including the beacon beam and the 4-quadrant detector. This will also reduce the system cost and the difficulties in \gls{FSO} calibration.

In this work, we demonstrate a beaconless and quadrant-detector-free alignment algorithm for \gls{HG}-mode-based \gls{MDM} system. By minimising power leakage to unwanted modes, this algorithm enables precise alignment to reduce inter-mode crosstalk. By eliminating redundant structures in alignment systems, it also reduces the overall cost of \gls{MDM} \gls{FSO} systems. When employing 6 lowest-order modes, this algorithm supported a 6 (modes)$\times$2 (polarisations)$\times$80\,GBaud 16-QAM transmission, obtaining a record-high line rate of 3.84\,Tbit/s in single-wavelength \gls{FSO} communication systems with a \gls{BER} below the \gls{HD-FEC} limit. By employing \gls{PS} 256-QAM in the same system, we achieved a record-high \gls{GMI} of 5.14 Tbit/s in single-wavelength \gls{FSO} communication systems.


\section{Results}\label{sec:re}
\subsection{Hermite-Gaussian modes and the local maximums}\label{sec:re.hg}

As an orthogonal complete set to decompose a coherent paraxial beam, the electric field of the \gls{HG} laser modes which propagate along the z-axis,  can be described as \cite{beijersbergen1993astigmatic}
\begin{equation}\label{equ:1}
\begin{aligned}
  {E_{l,m}}\left( {x,y,z} \right) & = C_{l,m}^{HG}\frac{1}{{{w_z}}}\exp \left( { - \frac{{{x^2} + {y^2}}}{{w_z^2}}} \right)\exp \left[ { - \frac{{ik\left( {{x^2} + {y^2}} \right)}}{{2{R_z}}}} \right] \\ 
   & \times {H_l}\left( {\frac{{\sqrt 2 x}}{{{w_z}}}} \right){H_m}\left( {\frac{{\sqrt 2 y}}{{{w_z}}}} \right)\exp \left( { - i{\psi _z}} \right)\exp \left( { - ikz} \right) ,\\ 
\end{aligned}
\end{equation}
where $C_{l,m}^{HG} = {\left( {{2^{l + m - 1}}\pi l!m!} \right)^{ - 1/2}}$ is the normalisation factor, ${R_z} = \left( {z_R^2 + {z^2}} \right)/z$ is the radius of curvature, ${z_R} = \pi w_0^2n/\lambda$ is the Rayleigh range, $w_0$ is the radius of beam waist, $n$ is the refractive index, $\lambda$ is the wavelength, ${w_z} = {w_0}\sqrt {1 + {{\left( {z/{z_R}} \right)}^2}}$ is the beam radius at $z$, $k=2 \pi / \lambda$ is the wave number, $\psi_z = \left( {l + m + 1} \right) \arctan \left( {z/{z_R}} \right)$ is the Gouy phase, ${H_n}\left( x \right) = {\left( { - 1} \right)^n}\exp \left( {{x^2}} \right)\frac{{{{\text{d}}^n}}}{{{\text{d}}{x^n}}}\exp \left( { - {x^2}} \right)$ is the Hermite polynomial of order $n$.

\begin{figure}[tb]
    \centering
    \includegraphics[width=0.9\textwidth]{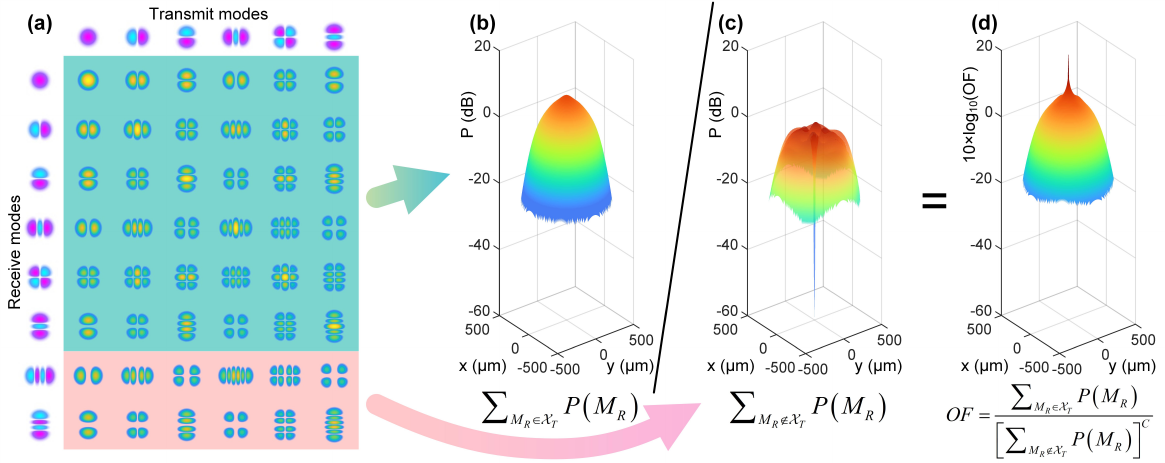}
    \caption{Concept of beaconless alignment. OF: objective function. (a) The scanning profile matrix from transmit \gls{HG} modes to receive \gls{HG} modes. (b) The received power summation of all transmitted modes at the receiver. (c) The received power summation of unwanted modes at the receiver. (d) The objective function.}
    \label{fig:align}
    \end{figure}

To calculate the received power from the transmitted mode $M_t = HG_{l_t,m_t}$ to the received mode $M_r = HG_{l_r,m_r}$, the equation below applies:
\begin{equation}\label{equ:2}
  P_{M_t,M_r} = {\left( {\iint {E_{{l_t},{m_t}}^*\left( {x - \Delta x,y - \Delta y,z} \right)}{E_{{l_r},{m_r}}}\left( {x,y,z} \right){\text{d}}x{\text{d}}y} \right)^2},
\end{equation}
where $\left( {\Delta x,\Delta y} \right)$ is the misalignment of the transmitted beam profile. As shown in Fig.~\ref{fig:align}(a), multiple local minimums occur when either the transmitted or the received mode is not $HG_{0,0}$, requesting the conventional beacon of an $HG_{0,0}$ (Gaussian) beam.

\subsection{Beaconless alignment}\label{sec:re.align}
To align the \gls{MDM} beam without a beacon, we need to find an objective function with only one local maximum where the system is optimally aligned.

\subsubsection{The requirement of transmit modes}\label{sec:re.hg.tx}
Intuitively, the overall power leakage to unwanted modes will increase when the misalignment is larger \cite{benton2024spatial}. Although multiple local maximums exist in Fig.~\ref{fig:align}(a), the power summation of all transmitted modes is a suitable objective function with only one local maximum (Fig.~\ref{fig:align}(b)), which can be written as:
\begin{equation}\label{equ:4}
  {P_{Tx}} = \sum\nolimits_{{M_r} \in {\mathcal{X}_T}} {P\left( {{M_r}} \right)},
\end{equation}
where $\mathcal{X}_T$ is the set of all transmitted modes, $P\left( {{M_r}} \right) = \sum\nolimits_{{M_t} \in {\mathcal{X}_T}} {{P_{{M_t},{M_r}}}}$ is the overall received power at receive mode $M_r$. Note the independence between the x- and y-axis indicated by Eq.~\eqref{equ:1}, we can set $y=0$ and analyse the power distribution in the x-axis without loss of generality. As shown in Fig.~\ref{fig:PowerProfile}(a), Eq.~\eqref{equ:4} works for all the tested situations from a total of 1 ($HG_{0,0}$) to 20 ($HG_{0,0} \sim HG_{19,0}$) transmitted modes. It is also worth noting that if $HG_{L,M}$ is chosen as a transmitted mode, all $HG_{l,m}$ modes that satisfy conditions $0\le l \le L$ and $0\le m \le M$ should also be chosen as transmitted modes to guarantee one single local maximum in Eq.~\eqref{equ:4}.

Although the transmit \gls{HG} modes form an orthogonal basis, their projections on the receive \gls{HG} modes are not necessarily orthogonal to each other when misalignment exists. When unmodulated laser beams with inherent slow phase shifts are loaded into different transmit modes, the power collected by a specific receive mode will exhibit fluctuations. These fluctuations are due to the interference effect, which is influenced by the phase variations of different transmit modes. To eliminate this fluctuation in beaconless \gls{FSO} systems, the phase of the laser source should be equally distributed in the time domain. This can be naturally achieved in coherent systems with phase modulation or \gls{QAM}, or in direct detection systems with a slight modification to the intensity modulators to introduce a random phase shift of $\pm \pi /2$.






\subsubsection{The requirement of receive modes}\label{sec:re.hg.rx}
As shown in Fig.~\ref{fig:PowerProfile}(a), the width of the ``flat" top will increase when the number of modes increases, leading to a worse alignment accuracy when noise and practical imperfection exist. Therefore, a ``sharp" metric is required when the system is nearly aligned.
Exploiting the intrinsic orthogonality in \gls{HG} modes, no power leakage to unwanted modes will be observed when the system is properly aligned (refer to the pink area in Fig.~\ref{fig:align}(a) for details). However, a small alignment error will introduce significant power leakage, especially to adjacent \gls{HG} modes (e.g. from $HG_{l,m-1}$ to $HG_{l,m}$ in Fig.~\ref{fig:align}(a)). As shown in Fig.~\ref{fig:align}(c), when the system is properly aligned, a sharp minimum can be observed in
\begin{equation}\label{equ:5}
  {P_{Rx}} = \sum\nolimits_{{M_r} \notin {\mathcal{X}_T}} {P\left( {{M_r}} \right)}.
\end{equation}
However, $P_{Rx}$ will also reduce when the beam goes beyond the aperture, preventing Eq.~\eqref{equ:5} from being a stand-alone objective function. Fortunately, a higher mode order leads to a larger effective area. Therefore, dividing Eq.~\eqref{equ:4} by Eq.~\eqref{equ:5} will remove the problem of multiple local extremums when the misalignment goes to infinity.
On the other hand, several local extremums are observed in Fig.~\ref{fig:align}(c) when the misalignment is a finite non-zero number. To eliminate these local extremums in the objective function, the monotonicity of Eq.~\eqref{equ:4} is exploited and a shaping coefficient $C \le 1$ is introduced in Eq.~\eqref{equ:3}.

Given the x-axis misalignment insensitivity of the unwanted mode $HG_{0,m+1}$ and y-axis misalignment insensitivity of the unwanted mode $HG_{l+1,0}$, it is suggested to incorporate at least two additional receive modes (e.g. $HG_{0,m+1}$ and $HG_{l+1,0}$) to improve alignment accuracy and obtain directional sensitivity, improving the acquisition speed in a quadrant-detector-free system. Moreover, including extra receive modes (e.g. $HG_{0,m+2}$ and $HG_{l+2,0}$) can further suppress practical errors and enhance the performance of the objective function. Importantly, it should be noted that in a quadrant-detector-free system, the directional sensitivity is unattainable in a Gaussian-beacon-based alignment system.

\subsubsection{The objective function}\label{sec:re.hg.of}
As shown in Fig.~\ref{fig:align}(d), invoking Eq.~\eqref{equ:4} and Eq.~\eqref{equ:5}, one suitable objective function can be written as 
\begin{equation}\label{equ:3}
  OF = \frac{{{P_{Tx}}}}{{P_{Rx}^C}} = \frac{{\sum\nolimits_{{M_r} \in {\mathcal{X}_T}} {P\left( {{M_r}} \right)} }}{{{{\left[ {\sum\nolimits_{{M_r} \notin {\mathcal{X}_T}} {P\left( {{M_r}} \right)} } \right]}^C}}},
\end{equation}
where $C \le 1$ is a shaping coefficient.
Multiple benefits can be obtained by employing the proposed objective function, including:
\begin{enumerate}
    \item
    \textbf{Better alignment accuracy:} When the system is nearly aligned, utilising the power leakage to the unwanted modes (Eq.~\eqref{equ:5}) yields a 'sharp' metric (Fig.~\ref{fig:align}(d)) in the proposed objective function (Eq.~\eqref{equ:3}), thereby achieving significantly higher accuracy compared to conventional methods. By excluding the beacon beam, the practical error between the signal beam and the beacon beam is also eliminated.

    \item
    \textbf{Faster acquisition:} The utilisation of transmitted higher-order modes (Eq.~\eqref{equ:4}) in the proposed objective function (Eq.~\eqref{equ:3}) allows a larger power profile (Fig.~\ref{fig:PowerProfile}(a)). Without sacrificing alignment accuracy, a larger acquisition step can be employed to reduce the number of iterations for faster acquisition.

    \item
    \textbf{Simplified system calibration:} The system calibration process is significantly simplified by excluding all the redundant optical structures such as beacon beam or 4-quadrant detector.

    \item
    \textbf{Better cost and energy efficiency:} The proposed objective function can be achieved by employing simple photodiodes, typically integrated at the input of receiver \glspl{EDFA} for amplification control. By excluding all the redundant optical structures, the system cost can be significantly reduced, with a higher energy efficiency by excluding power absorption on the redundant devices (e.g. wavelength beam splitters).
\end{enumerate}





\subsection{Experimental setup}\label{sec:re.es}
    \begin{figure}[tb]
    \centering
    \includegraphics[width=0.9\textwidth]{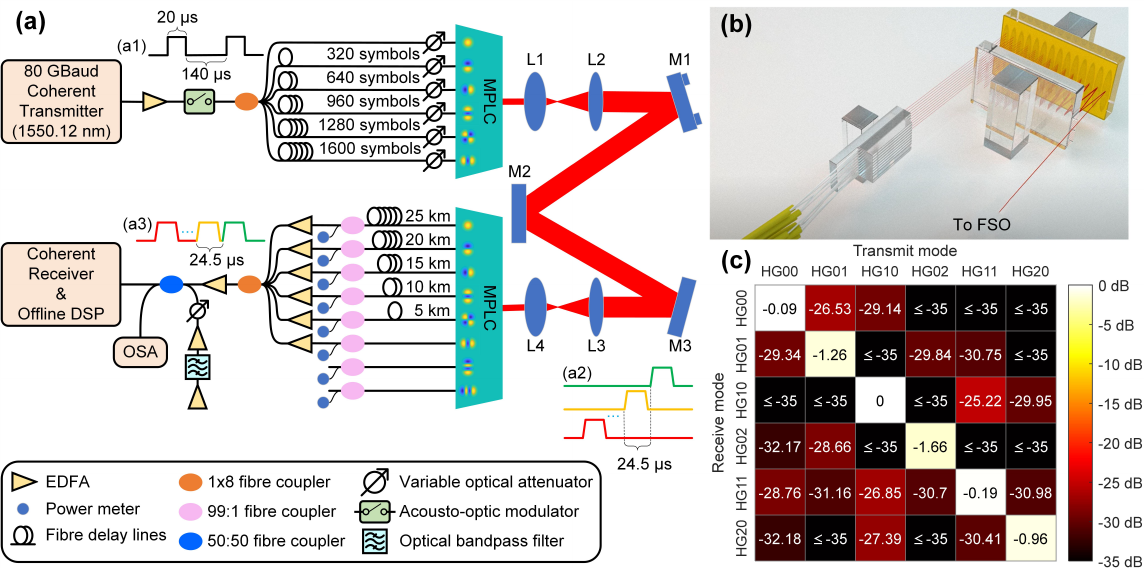}
    \caption{Experimental setup for beaconless \gls{MDM} communication system. L: lens; M: mirror; EDFA: erbium-doped fiber amplifier; MPLC: multi-plane light conversion; OSA: optical spectrum analyser; FSO: free-space optics. (a) Schematical diagram of experimental setup. (b) The structure of the \gls{MPLC} device. (c) The normalised transfer matrix when the system is properly aligned.}
    \label{fig:system}
    \end{figure}
Our experimental \gls{MDM} \gls{MIMO} \gls{FSO} communication system is depicted in Fig.~\ref{fig:system}. At the transmitter, a 1550.12\,nm laser was modulated by a \gls{DP} \gls{IQ} modulator with a nominal 6\,dB bandwidth of 45\,GHz using a 4-channel 120\,GSa/s \gls{AWG} with a 3\,dB bandwidth of 50\,GHz, and a 80\,GBaud signal was generated. The signal was \gls{RRC} shaped with a roll-off factor of 0.1. The signal had a frame structure of 20,000 symbols with 1,920 \gls{QPSK} symbols in the training sequence, and 1 random \gls{QPSK} pilot symbol for every 9 data symbols, the data symbols were either 16-QAM symbols generated from a $2^{15}-1$ \gls{PRBS} or PS-256-QAM symbols generated from a mt19937ar Mersenne Twister \cite{matsumoto1998mersenne}. After amplified by an \gls{EDFA}, the signals were passed through an \gls{AOM} to generate a 20\,\textmu s burst signal for each 160\,\textmu s period (Fig.~\ref{fig:system}(a1)) to enable the \gls{TDM} receiver setup which will be detailed later. To emulate independent transmitters, 6 copies were split from the burst signal and delayed by variable \glspl{FDL} with lengths of 0, 320, 640, 960, 1280, and 1600 symbols. The delayed signals were passed through \glspl{VOA} to compensate for practical power differences before the \glspl{VOA}. The compensated signals were then converted to corresponding \gls{HG} modes using \gls{MPLC} technology (Fig.~\ref{fig:system}(b), the output beam waist diameter of $HG_{0,0}$ mode is approximately 0.3\,mm) \cite{fontaine2019laguerre}.

In the \gls{FSO} channel, the transmitted \gls{HG} modes were passed through a beam expander consisting of a convex lens with a focal length of 30\,mm (L1) followed by another convex lens with a focal length of 100\,mm (L2). The expanded beam was passed through a steering mirror (M1) with two high-precision motorised actuators for automatic alignment. The stepper motors of the actuators support an incremental motion of approximately 1\,{\textmu}m in standard mode or approximately 0.5\,nm using micro-steps in high-precision mode, providing adequate precision for our experiments. The beam was then passed through two more mirrors (M2, M3), and another symmetric beam expander consisting of a convex lens with a focal length of 100\,mm (L3) followed by another convex lens with a focal length of 30\,mm (L4). The beam was then coupled into the receiver \gls{MPLC} device after a total \gls{FSO} length of \textasciitilde1.6\,m. 
As shown in Fig.~\ref{fig:system}(c), the normalised transfer matrix can achieve $<$1.7\,dB mode-dependent loss, and $>$25\,dB suppression for $6\times6$ \gls{MIMO} when the system is properly aligned using the proposed algorithm.

At the receiver, the optical signals split by the receiver \gls{MPLC} device were delayed by the \glspl{FDL} for the 6 lowest order modes to realise the \gls{TDM} receiver. Here, a 5\,km difference between adjacent modes introduced a \textasciitilde24.5\,\textmu s delay (Fig.~\ref{fig:system}(a2)), slightly longer than the signal burst (Fig.~\ref{fig:system}(a1)). Each delayed signal was then split by a 99:1 fibre coupler so that the overall power received by each mode could be measured by a corresponding power meter. Here, we also connected two more unused higher-order modes (i.e. $HG_{0,3}$ and $HG_{3,0}$) to calculate the denominator term in Eq.~\eqref{equ:3}. The 6 lowest order modes were then amplified by 6 \glspl{EDFA}, coupled into one \gls{SMF} for \gls{TDM} combining  (Fig.~\ref{fig:system}(a3)), and amplified by another \gls{EDFA}. 
Afterwards, the variable optical noise was loaded by a sequence of devices involving an \gls{EDFA}, an optical bandpass filter, another \gls{EDFA}, a \gls{VOA}, and a 50:50 coupler. The average \gls{OSNR} was then accessed using an \gls{OSA} to facilitate OSNR scanning.
Finally, the \gls{TDM} signals were received by a coherent receiver with a 70\,GHz, 200\,GSa/s oscilloscope and demodulated by offline \gls{DSP}.

\subsection{Experimental results on alignments}\label{sec:re.hg.exp}
    \begin{figure}[tb]
    \centering
    \includegraphics[width=0.9\textwidth]{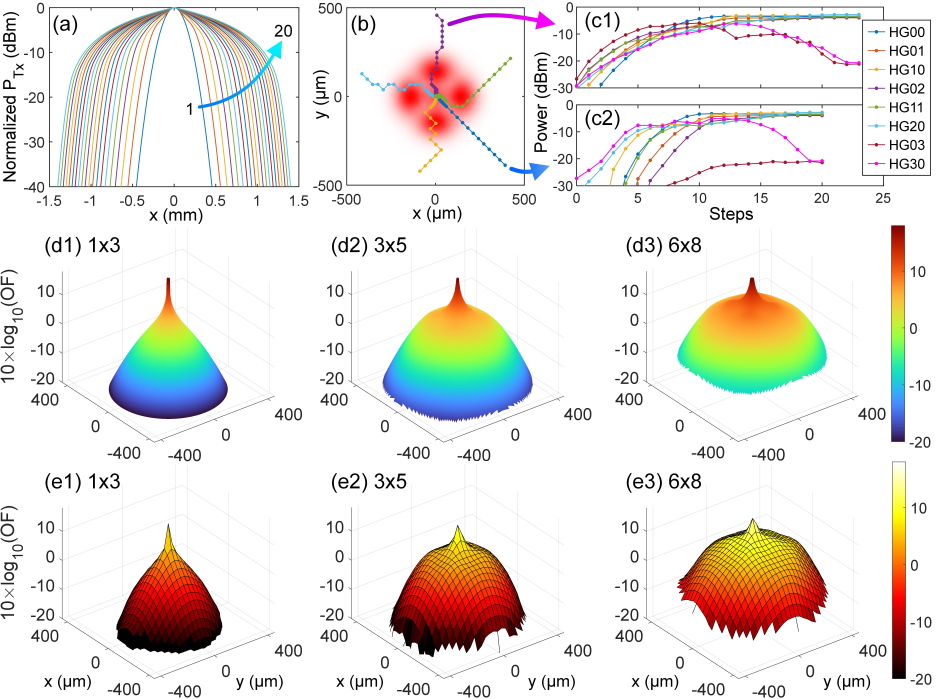}
    \caption{Summation of received power versus misalignment. (a) Received power profile versus x-axis misalignment. (b) Traces of alignments from different initial points. (c) Mode-wise received power versus alignment steps for the y-axis and diagonal misalignments, respectively. (d) Simulational objective function versus misalignment for $1\times3$, $3\times5$, and $6\times8$ systems, respectively. (e) Experimental objective function versus misalignment for $1\times3$, $3\times5$, and $6\times8$ systems, respectively.}
    \label{fig:PowerProfile}
    \end{figure}
We transmitted the 6 lowest order modes and employed $HG_{0,3}$ and $HG_{0,3}$ as unwanted receive modes for better alignment accuracy. Equipped with the proposed objective function, we're ready to align the system by employing the greedy algorithm \cite{fernandes2021adaptive}. Our strategy was to sequentially move towards nine neighbouring points including $\left( {-\Delta x,-\Delta x} \right)$, $\left( {-\Delta x,0} \right)$, $\left( {-\Delta x,\Delta x} \right)$, $\left( {0,-\Delta x} \right)$, $\left( {0,0} \right)$, $\left( {0,\Delta x} \right)$, $\left( {\Delta x,-\Delta x} \right)$, $\left( {\Delta x,0} \right)$, and $\left( {\Delta x,\Delta x} \right)$, where $\Delta x$ is the step size. We recorded the objective function for all nine points and specifically targeted the point with the maximum value of the objective function. Repeating the greedy algorithm ensures precise alignment by focusing on the most favourable point within the immediate vicinity. The alignment traces of a $6\times8$ \gls{MDM} system from different starting points are depicted in Fig.~\ref{fig:PowerProfile}(b). A beam profile including 6 transmitted modes
is also depicted in Fig.~\ref{fig:PowerProfile}(b) as a reference. Here the HG modes are rotated by $45^\circ$ for symmetric considerations when designing the \gls{MPLC} device \cite{fontaine2019laguerre}. Despite the choice of different initialisation points, the system could always converge to the centre of the beam profile. Moreover, the alignment traces tend to follow diagonal directions, which aligns with the “diagonal ridge” effect in the objective function as shown in Fig.~\ref{fig:PowerProfile}(d3) and Fig.~\ref{fig:PowerProfile}(e3).

The mode-wise received power of two typical alignment curves (along the y-axis and diagonal direction, respectively) is depicted in Fig.~\ref{fig:PowerProfile}(c1)-(c2). As anticipated, the power of transmitted modes reached its maximum, while the power of unwanted modes was minimised when the system was properly aligned. After step 15, the unwanted modes made substantial contributions when the power of the transmitted mode reached a near-consistent level, indicating the importance of the ``sharp'' metric in the proposed objective function for precise alignment. Furthermore, the directional sensitivity between $HG_{l,0}$ and $HG_{0,m}$ is evident in Fig.~\ref{fig:PowerProfile}(c2) when the path follows the same direction as $HG_{l,0}$ modes.

By deliberately misaligning the system away from its aligned central point, we can compare the experimental objective function against the simulation results. As shown in Fig.~\ref{fig:PowerProfile}(d1)-(d3), the objective function works for $1\times3$, $3\times5$, and $6\times8$ \gls{MIMO} systems when the lowest order modes (i.e. smallest $l+m$) were employed. The experimental results in Fig.~\ref{fig:PowerProfile}(e1)-(e3) align well with the corresponding simulation predictions. Moreover, the symmetry in the experimental results also indicated the high precision of the proposed objective function.

\subsection{Achieved BER and data throughput}

    \begin{figure}[tb]
    \centering
    \includegraphics[width=0.9\textwidth]{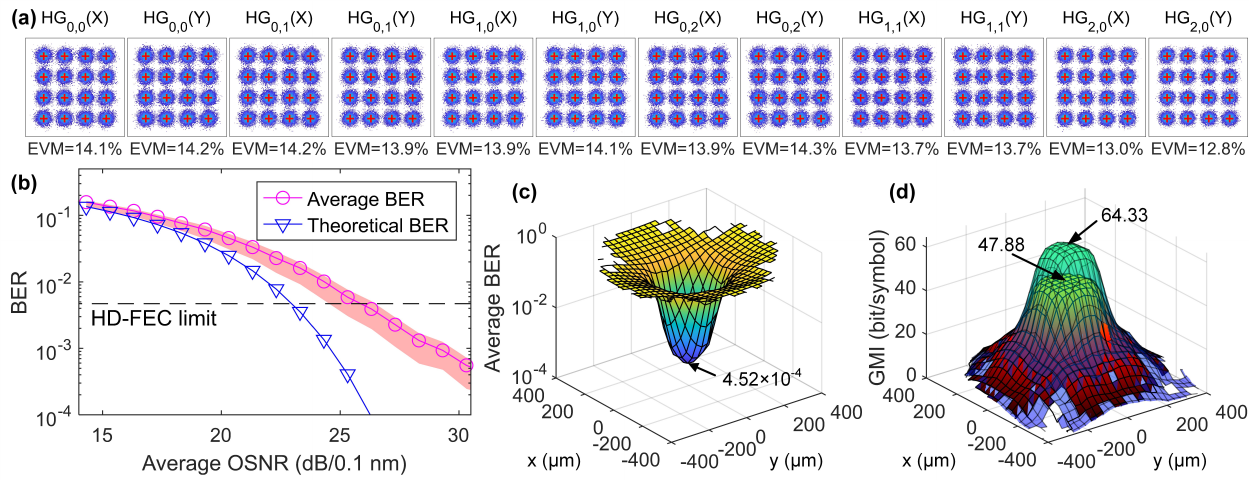}
    \caption{Experimental performance of communication system. EVM: error vector magnitude; BER: bit error rate; OSNR: optical signal-to-noise ratio; HD-FEC: hard-decision forward error correction; GMI: generalized mutual information. (a) Constellation diagram. (b) BER versus average OSNR. (c) BER versus misalignment. (d) GMI versus misalignment.}
    \label{fig:ber}
    \end{figure}
Fig.~\ref{fig:ber} shows the experimental performance of the proposed communication system, including the constellation diagram and the \gls{BER} for 16-QAM signals, and the achievable data rate estimated from \gls{GMI} for 16-QAM and PS-256-QAM signals, respectively.
Fig.~\ref{fig:ber}(a) depicts the channel-wise constellation diagrams for 16-QAM signals. Benefit from the proposed alignment algorithm, the constellations perform similarly in all channels, while some differences in \gls{EVM} are observed due to the mode-dependent loss introduced by the \gls{MPLC} devices.
Fig.~\ref{fig:ber}(b) shows the mode-wise \gls{BER} versus \gls{OSNR} for 16-QAM signals. Here we observe an implementation penalty of approximately 2.9\,dB from the theoretical limit at the \gls{HD-FEC} limit of $4.7\times10^{-3}$ \cite{alvarado2015replacing}. This is mainly due to practical implementation loss including shot noise and quantisation noise introduced by the digital coherent receiver. As illustrated by the pink area, the channel-wise \gls{BER} shows slight variations. Consistent with the \gls{EVM} difference shown in Fig.~\ref{fig:ber}(a), these variations are primarily attributed to the mode-dependent loss in the \gls{MPLC} devices.

Fig.~\ref{fig:ber}(c) depicts the \gls{BER} performance versus the misalignment when the \gls{OSNR} loading was disabled. When the system was properly aligned, we obtained a \gls{BER} of $4.52\times10^{-4}$, an order of magnitude smaller than the \gls{HD-FEC} limit. Moreover, an approximately 34\% \gls{BER} degradation was observed when the beam was deliberately misaligned by 30\,{\textmu}m, which was 1/10 of the $HG_{0,0}$ beam diameter. Furthermore, the degradation significantly increased as the misalignment increased. This phenomenon indicated the importance of accurate alignment provided by our proposed algorithm.
Fig.~\ref{fig:ber}(d) depicts the \gls{GMI} performance versus misalignment when the \gls{OSNR} loading was disabled. The maximum \gls{GMI} of 16-QAM signals are 47.88\,bit/symbol, slightly smaller than the theoretical limit of 6 (modes)$\times$2 (polarisations)$\times$4\,bit/symbol. By loading the PS-256-QAM signals, we obtained a maximum \gls{GMI} of 64.33\,bit/symbol, equivalent to an achievable line rate of 5.146\,Tbit/s when transmitting 80\,GBaud signals.

\section{Methods}\label{sec:me}

\subsection{DSP details}\label{sec:me.dsp}

A nonlinear Volterra predistorter was employed to compensate for the nonlinearities and imperfections in the 80\,GBaud system \cite{eun1997new,nolle2020characterization}. To perform the system identification, an 80\,GBaud DP-16QAM probe signal with a roll-off factor of 0.1 was employed and a nonlinear filter with kernel memory lengths of (201,9,9) taps was created.

The offline \gls{DSP} includes a sequential combination of the Gram-Schmidt orthogonalisation \cite{fatadin2008compensation}, polarisation demultiplexing in Stokes space \cite{szafraniec2010polarization}, frequency-domain frequency offset estimation, frequency-domain \gls{CD} compensation for the receiver \glspl{FDL} \cite{faruk2017digital}, Godard timing recovery \cite{godard1978passband}, phase-asynchronous phase and channel estimation \cite{li2024phase}, and 3-dimensional \gls{MIMO} equalisation \cite{randel20116}.

The shaping factor of PS-256-QAM is set to 2 to obtain $<$0.1\,dB \gls{SNR} penalty when estimating data rate from \gls{GMI} for up to 6\,bit/symbol for the \gls{QAM} constellations \cite{fehenberger2016probabilistic}.

\subsection{Measurement of transfer matrix}\label{sec:me.MoTM}

The normalised transfer matrix given in Fig.~\ref{fig:system}(c) was measured by sequentially connecting each mode of the transmitter \gls{MPLC} directly to the transmitted signal which was amplified to 10\,dBm by an \gls{EDFA}. The power attenuation introduced by receiver \glspl{FDL} were compensated to reflect proper power readings directly after the receiver \gls{MPLC}. The power measurement of $\le$-35\,dBm in Fig.~\ref{fig:system}(c) was due to the limited sensitivity of the power meters.

\section{Discussion}\label{sec12}

We proposed a novel objective function for beaconless and quadrant-detector-free \gls{MDM} \gls{FSO} auto-alignment. This method significantly improved the alignment accuracy and acquisition speed. It also eliminated redundant hardware, reducing system cost, improving energy efficiency and facilitating system calibration. Moreover, this method supports a complete modal basis of Hermite-Gaussian modes. 
With identical beam divergence, this will enable the transmission of a significantly larger number of spatial channels compared to the incomplete OAM mode basis.

Our \gls{MDM} \gls{FSO} communication experiments demonstrated that the proposed alignment algorithm enables an inter-mode crosstalk suppression of $>$25\,dB in our $6\times6$ \gls{MDM} \gls{FSO} communication system. By employing Volterra-based system identification, we achieved a record-high 80\,GBaud in \gls{MDM} transmissions, supporting up to a record-high 3.84\,Tbit/s and 5.14\,Tbit/s line rate when employing 16-QAM and PS-256-QAM signals, respectively. 

Our proposed algorithm has shown superb alignment accuracy, inter-mode crosstalk suppression, and hardware efficiency. By exploiting spatial diversity, our experimental demonstration underpins the potential of ultra-high data rate \gls{MDM} transmission. We anticipate this paper to be a starting point for more sophisticated alignment scenarios in future \gls{MDM} \gls{FSO} communications for long-distance applications with generalised mode basis such as \gls{LG} modes, Ince-Gaussian modes, Bessel modes, etc.

\backmatter





\bmhead{Acknowledgements}
We acknowledge the support by EPSRC under Grants EP/T009047/1, EP/S003436/1, and EP/S016171/1, and by the European Union’s Horizon 2020 research and innovation programme under the Marie Skłodowska-Curie Grant 713694.

\bmhead{Author contributions}
Y.L. and G.F. designed and carried out the experiment and analysed the data. A.B. designed and assembled the multi-plane light conversion devices. Y.L. and D.B. designed and calibrated the optical system. Y.L. designed and calibrated the DSP. M.P. and A.E. provided the technical support. Y.L. drafted the manuscript with support from all co-authors. A.E. supervised the project.

\bmhead{Competing interests} 
The authors declare no competing interests.










\bibliography{sn-bibliography}

\end{document}